# Flow-Induced Deformation of a Flexible Thin Structure as Manifestation of Heat Transfer Enhancement

Atul Kumar Soti[1], Rajneesh Bhardwaj[2*] and John Sheridan[3]

[1]IITB-Monash Research Academy, Indian Institute of Technology Bombay, Mumbai, 400076, India

[2]Department of Mechanical Engineering, Indian Institute of Technology Bombay, Mumbai, 400076 India

[3]Department of Mechanical &Aerospace Engineering, Monash University, Clayton, VIC 3800, Australia

[*]Corresponding author (rajneesh.bhardwaj@iitb.ac.in)

Phone: +91 22 2576 7534, Fax: +91 22 2572 6875

*Abstract*

Flow-induced deformation of thin structures coupled with convective heat transfer has potential applications in energy harvesting and is important for understanding functioning of several biological systems. We numerically demonstrate large-scale flow-induced deformation as an effective passive heat transfer enhancement technique. An in-house, strongly-coupled fluid-structure interaction (FSI) solver is employed in which flow and structure solvers are based on sharp-interface immersed boundary and finite element method, respectively. In the present work, we validate convective heat transfer module of the in-house FSI solver against several benchmark examples of conduction and convective heat transfer including moving structure boundaries. The thermal augmentation is investigated as well as quantified for the flow-induced deformation of an elastic thin plate attached to lee side of a rigid cylinder in a heated channel laminar flow. We show that the wake vortices past the plate sweep higher sources of vorticity generated on the channel walls out into the high velocity regions – promoting the mixing of the fluid. The self-sustained motion of the plate assists in convective mixing, augmenting convection in bulk and near the walls; and thereby reducing thermal boundary layer thickness as well as improving Nusselt number at the channel walls. We quantify the thermal improvement with respect to channel flow without any bluff body and analyze the role of Reynolds number, Prandtl number and material properties of the plate in the thermal augmentation.

# 1 Introduction

Fluid-structure interaction (refereed as FSI hereafter) of flexible thin structures immersed in a flow coupled with convective heat transfer enhances thermal transport, which could be utilized to improve the thermal performance of engineering as well as biological systems. Examples include, energy harvesting [1], microelectronics cooling and microchannels, heat exchangers. Examples of biological systems include, thermoregulation in elephants via flapping of their large ears [2] and thermal transport enhancement in microchannels using oscillating synthetic cilia [3].

While analyzing the FSI of flexible thin structures without heat transfer, several previous numerical studies ignored internal stresses and thickness of the structure, and considered it as a membrane (see review by Shelley and Zhang [4]). Zheng et al. [5] considered the internal stresses in the structure in their FSI model; however, they considered small-scale deformation of the structure and neglected geometric non-linearity in the structure solver. In the context of large-scale deformation, Baaijens [6] presented numerical analysis of the FSI of slender bodies placed in a channel flow; however, this study ignored inertial terms in the governing equation of the structure domain. Similarly, Vigmostad et al. [7] showed capability of simulating thin, flexible structures using a sharp-interface Cartesian grid method. A FSI benchmark involving large-scale flow-induced deformation was first proposed by Turek and Hron [8]. In this benchmark, an elastic plate attached to the lee side of a rigid cylinder develops self-sustained oscillation in 2D laminar channel flow. The FSI studies of flexible thin structures by Dunne and Rannacher [9], Heil et al. [10], Bhardwaj and Mittal [11], Lee and You [12] and Tian et al. [13] showed validation of their respective solvers against the benchmark proposed by Turek and Hron [8].

The other class of FSI numerical studies considered coupled convective heat transfer; however, these studies ignored the flexibility of the immersed structure and considered it rigid. These studies demonstrated heat transfer enhancement and can be conveniently categorized in passive and active techniques [14, 15], which either harness flow energy or utilize external source of energy to augment the heat transfer, respectively. The flow past bluff, rigid bodies such as, rectangular cylinder [16], inclined square cylinder [17], triangular cylinder [18], wings [19, 20, 21] in a channel, conical [22] and louvered strip inserts [23] in a tube, and flow in wavy-wall channel [24] were utilized as a passive technique. On the other hand, the heat transfer improvement via oscillating bluff bodies inside the channel [25, 26] was shown as the active technique. Recently, Mills et al. [3] considered an array of oscillating synthetic cilia to show heat transfer enhancement in microchannels, however, they did not consider flow-induced deformation in their model.

Very few numerical studies which considered FSI of the flexible structures coupled with convective heat transfer were reported. For instance, Khanafer et al. [27] simulated a heated flexible cantilever attached to a square cylinder in a channel. Very recently, Soti and Bhardwaj [28] demonstrated heat transfer enhancement via large-scale flow-induced deformation in the FSI benchmark at Reynolds number, $Re$ = 100. They compared effectiveness of various configurations obtained from FSI benchmark using top-down approach (Table 1). Similarly, Shi et al. [29] showed thermal improvement in the FSI benchmark at $Re$ = 204.8 - 327.7.

In addition, there is limited availability of numerical FSI benchmarks involving large-scale deformation coupled with convective heat transfer. In this context, the objective of the present work is twofold: First, numerically demonstrate and quantify large-scale flow-induced deformation as an effective passive heat transfer enhancement technique. Second, we provide benchmark results which will serve as validation data for FSI solvers coupled with convective heat transfer. Since several previous numerical studies [9, 10, 11, 12, 13] used FSI benchmark proposed by Turek and Hron [8] to validate FSI solvers, we propose an extension of this FSI benchmark to account for coupled convective heat transfer in the present paper.

In present work, we built upon our previous work [28] and employ a sharp-interface immersed boundary method based flow and heat transfer solver with finite-element based structural dynamics solver (section 2). First, we perform code validation for the heat transfer module of the in-house FSI solver (section 2.1). Second, we consider convective heat transfer in FSI benchmark problem proposed by Turek and Hron [8] and explain role of vortex dynamics in enhanced mixing of the fluid in section 3.1. The mechanism of thermal augmentation via large-scale flow-induced deformation is described in section 3.2. The quantification of the heat transfer improvement is analyzed by calculating and comparing the effectiveness of various configurations (Table 1) with respect to pumping power required in the channel in section 3.3. Finally, we quantify the influence of several relevant parameters such as Prandtl number, Reynolds number, Young's Modulus and channel length in section 3.3.

## 2   Computational Model

An in-house FSI solver based on sharp-interface immersed boundary method is employed to simulate the fluid dynamics, heat transfer and structure dynamics. The governing equations of the flow domain are solved on a fixed Cartesian (Eulerian) grid while the movement of the immersed structure surfaces is tracked in Lagrangian framework. As reviewed by Mittal and Iaccarino [30], the immersed boundary method is relevant for 3D complex moving boundaries on a Cartesian grid. Since governing equations are solved on body non-conformal, Cartesian grid, there is no need of re-meshing with deforming or moving immersed structures in the fluid domain. The FSI solver employed in the present study was developed by Mittal and co-workers [5, 31, 32, 33] and later developed for large-scale flow-induced deformation by Bhardwaj and Mittal [11]. The flow is governed by unsteady, viscous and incompressible Navier-Stokes equations:

$$\frac{\partial v_i}{\partial x_i} = 0 \tag{1}$$

$$\frac{\partial v_i}{\partial t} + \frac{\partial v_i v_j}{\partial x_j} = -\frac{\partial p}{\partial x_i} + \frac{1}{Re}\frac{\partial^2 v_i}{\partial x_j^2} \tag{2}$$

where $i, j = 1, 2, 3$, and $v_i$, $t$, $p$ and $Re$ are velocity components, time, pressure and Reynolds number, respectively. These equations are discretized in space using a cell-centered, collocated (non-staggered) arrangement of primitive variables $v_i$, $p$ and a second-order, central-difference scheme is used for all spatial derivatives. In addition, the face center velocities are computed [34], which results in discrete mass conservation to machine accuracy. The implementation details can be found in previous and recent papers by Mittal and co-workers [5, 31, 32, 33]. Here we provide the methodology briefly. The unsteady Navier-Stokes equation is marched in time using a fractional-step scheme [31, 35] which involves two steps: solving an advection-diffusion equation followed by a pressure Poisson equation. During the first step, both the convective and viscous terms are treated implicitly using Crank-Nicolson scheme to improve the numerical stability. In the second step, the pressure Poisson equation is solved with the constraint that the final velocity be divergence-free. Once the pressure is obtained, the velocity field is updated to its final value in the final sub-step. A fast geometric multigrid solver [36] is used to solve the pressure Poisson equation. A sharp-interface immersed boundary method based on multi-dimensional ghost-cell methodology [31] is utilized and the immersed structure boundary is represented using unstructured grid with triangular elements in Cartesian volume grid of the flow domain.

The structure dynamics was simulated using an open-source finite-element solver Tahoe© [37]. The governing equations for the structure are the Navier equations (momentum balance equation in Lagrangian form) and are written in non-dimensional form as:

$$\rho_s \frac{\partial^2 d_i}{\partial t^2} = \frac{\partial \sigma_{ij}}{\partial x_j} + \rho_s f_i, \tag{3}$$

where $i$ and $j$ range from 1 to 3, $\rho_s$ is the dimensionless density of the structure with respect to the fluid density, $d_i$ is the displacement component in the $i$ direction, $t$ is the time, $\sigma$ is the Cauchy stress tensor and $f_i$ is the body force component in the $i$ direction. The structure solver was implicitly (two-way) coupled with the flow solver using a partitioned (segregated) approach by Bhardwaj and Mittal [11]. The implicit coupling ensures numerical stability at low structure-fluid density ratio [5]. The implementation details of the coupling can be found in Refs. [11, 38]. The structural solver Tahoe© [37] was coupled with a compressible flow solver in Ref. [38]. No slip boundary conditions are applied for the velocity at the fluid-structure interface which represents continuity of the velocity at the interface,

$$v_{i,f} = \dot{d}_{i,s}, \tag{4}$$

where subscripts $f$ and $s$ denote fluid and structure, respectively. In addition, the balance of forces is applied at the interface,

$$\sigma_{ij,f} n_j = \sigma_{ij,s} n_j, \tag{5}$$

where $n_j$ is the local surface normal pointing outward from the surface. The pressure loading on the structure surface exposed to the fluid domain is calculated at the current location of the structure using interpolated fluid pressure via bilinear interpolation, as described by Mittal et al. [31].

The heat transfer inside the fluid is governed by the following dimensionless energy equation:

$$\frac{\partial T}{\partial t} + v_i \frac{\partial T}{\partial x_i} = \frac{1}{Pe} \frac{\partial^2 T}{\partial x_j^2}, \tag{6}$$

where $Pe$ is Peclet number and $T$ is dimensionless temperature, defined in terms of dimensional temperature $T^*$, reference wall temperature $T^*_w$ and reference temperature $T^*_{\text{ref}}$, as follows,

$$T = \frac{T^* - T^*_{ref}}{T^*_w - T^*_{ref}}, \tag{7}$$

$T^*_{ref}$ is taken as inlet temperature of the fluid in all simulations unless specified otherwise. The thermal boundary conditions are described as follows (Fig. 1). The temperature is considered as uniform at the inlet. The fluid-structure interface is insulated and channel walls are at constant temperature. Zero Neumann temperature boundary conditions are applied at the outlet. The heat transfer augmentation is characterized using instantaneous Nusselt numbers at the channel wall, which is defined as follows [39]:

$$Nu(x,t) = \frac{2H}{T_m - 1} \left.\frac{\partial T}{\partial y}\right|_{wall}, \tag{8}$$

where $2H$ is the dimensionless hydraulic diameter of the channel. The bulk mean temperature, $T_m$, is defined as [39]:

$$T_m(x,t) = \frac{\int_0^H uTdy}{\int_0^H udy}, \tag{9}$$

where $u$ stands for dimensionless axial velocity component. Time-average Nusselt number ($Nu_{avg}$) over one period of oscillation, $\tau$, for time-periodic flow is expressed as:

$$Nu_{avg}(x) = \frac{1}{\tau}\int_t^{t+\tau} Nu(x,t)dt, \tag{10}$$

Time- and space-average Nusselt number ($Nu_{mean}$) over surface area $A$ is defined as:

$$Nu_{mean} = \frac{1}{A}\iint_A Nu_{avg}(x)dA \tag{11}$$

## 2.1 Code validation

The flow solver was extensively validated by Mittal et al. [31] against several benchmark problems such as flow past a circular cylinder, sphere, airfoil and suddenly accelerated circular cylinder and normal plate and was used to simulate 3D biological flows involving FSI [5, 33]. The capability of simulating large-scale flow-induced deformation with implicit (two-way) coupling in the FSI solver was implemented by Bhardwaj and Mittal [11]. We briefly present validation results for this module in section 2.1.1 for the sake of completeness of the validation of the model. The validation studies carried out in the present work for convective heat transfer module are described in subsequent subsections.

### 2.1.1 Large-scale flow-induced deformation

The flow-induced deformation module in the in-house FSI solver was validated by Bhardwaj and Mittal [11] against the FSI benchmark proposed by Turek and Hron [8]. In the benchmark, a 3.5$D$ × 0.2$D$ elastic plate attached on the lee side of a rigid cylinder is placed inside a channel of width 4.1$D$, where $D$ is the cylinder diameter (Fig. 1). The fluid is considered to be Newtonian and incompressible. The plate is considered to be of Saint Venant-Kirchhoff material, which accounts for geometric-nonlinearity for a linear elastic material [40]. The boundary conditions for the benchmark problem are shown in Fig. 1. At the inlet, a fully developed parabolic velocity profile with mean velocity $U_m = 1$ is applied and no slip boundary conditions are applied at the channel walls and immersed structure boundaries. At the outlet, zero Neumann boundary condition is applied for the velocity. Note that the thermal boundary conditions shown in Fig. 1 will be used later in the discussion. The Reynolds number, dimensionless Young's modulus, structure-fluid density ratio and Poisson's ratio are 100, $1.4 \times 10^3$, 10 and 0.4, respectively. The minimum grid sizes in $x$ and $y$ directions are $\Delta x_{min} = 0.0231$ and $\Delta y_{min} = 0.02$, respectively, and time-step is $\Delta t = 0.01$. The validation in Ref. [11] was conducted for time-varying position of the tip of the plate (Fig. 2) and oscillation frequency, after the plate reaches self-sustained periodic oscillation state. The position as well as frequency was in good agreement with published results of Turek and Hron [8].

### 2.1.2 Conduction heat transfer

We present validation of 1D unsteady and 2D steady heat conduction cases against analytical model [41]. The temperature in these cases is governed by the heat conduction equation,

$$\frac{\partial T^*}{\partial t^*} = \alpha \frac{\partial^2 T^*}{\partial x_j^{*2}}, \tag{12}$$

where $\alpha$ is thermal diffusivity of the material. The dimensionless form of eq. (12) with a reference velocity and length renders eq. (6) with flow velocity set to zero. The reference velocity in both cases is taken as 1 m/s. In 1D unsteady case, the reference length, the reference wall temperature ($T^*_w$) and reference temperature ($T^*_{ref}$) are taken as the slab length, right wall temperature and initial slab temperature, respectively. The left as well as right boundary is at $T_w = 1$ (Fig. 3a). The grid-size and time-step are $1.5 \times 10^{-2}$ and $1 \times 10^{-4}$, respectively. In Fig. 3b and 3c, we note good agreement between numerical and analytical results at different time instances for $Pe = 1$ and $Pe = 4$, respectively.

In 2D steady-state case, a square block of dimension $1 \times 1$ with a circular hole of diameter $D = 0.5$ at its center (Fig. 4(a)) is considered with $Pe = 1$ and $Re = 1$. The reference length, reference wall temperature ($T^*_w$) and reference temperature ($T^*_{ref}$) are taken as the block size, right wall temperature and left wall temperature, respectively. The left and right boundaries of the block are kept at temperatures 0 and 1, respectively with top and bottom boundaries insulated. A uniform heat flux $q''_w = 1$ is applied along the boundary of the hole. The unsteady heat conduction equation (eq. (6)) was solved with uniform grid-size and time-step $1.5 \times 10^{-2}$ and $1 \times 10^{-3}$, respectively, until steady state is reached. In Fig. 4(c), we plot the temperature variation along the x-axis at $y = 0.5$, calculated by our numerical simulation and MATLAB partial differential equation solver [42]. The comparison is in good agreement and verifies our calculations. A qualitative comparison of isotherms is shown in Fig. 4(a) and (b).

### 2.1.3 Convective heat transfer without bluff body

We validate convective heat transfer module with analytical model given in Ref. [39] for steady-state channel flow with constant wall temperature and constant wall heat flux, respectively. In this validation, we consider a $20H \times H$ channel and dimensional reference temperature $T^*_{ref}$ is taken as wall temperature. At inlet, a fully-developed parabolic velocity is applied with maximum value of $1.5U_m$ at the center, $U_m$ being the mean velocity. Zero Neumann velocity boundary condition is applied at outlet. The Reynolds number, based on the mean velocity and channel hydraulic diameter ($2H^*$), and Prandtl number are taken as 200 and 1, respectively. A uniform inlet temperature, $T_{in} = 1$ and 0 is considered for constant wall temperature ($T_w = 0$) and constant wall heat flux ($q''_w = 1$) cases, respectively. Zero Neumann boundary condition is applied for temperature at the outlet. Simulation is performed till steady state on a $512 \times 64$ uniform grid with $\Delta x_{min} = 0.04$ and $\Delta y_{min} = 0.02$ with time step equal to 0.02. Local Nusselt numbers ($Nu$) (eq. (8)) calculated using the present numerical model and analytical expressions (eq. 3.111-3.114 in Ref. [39]) are compared in Fig. 5. The fully developed values of $Nu = 7.54$ and 8.23 are obtained for the constant wall temperature and constant wall heat flux cases, respectively. These values as well as space-variation of *Nu* match well with the analytical theory and verify the convective heat transfer module.

### 2.1.4 Convective heat transfer with stationary immersed boundary

We validate convective heat transfer module with an immersed stationary boundary for free stream flow around a heated cylinder (diameter $D = 1$) in a $40D \times 20D$ channel. The computational domain and boundary conditions are shown in Fig. 6a. The simulations are performed for a range of Reynolds number. The Reynolds number (*Re*) based on free stream velocity and the cylinder diameter are varied in range of 80-200, keeping Prandtl number constant, $Pr = 0.7$. We utilize a 384×384 non-uniform grid with $\Delta x_{min} = 0.005$ and $\Delta y_{min} = 0.005$, and $\Delta t = 0.0025$. The simulated local Nusselt number for $Re = 120$ is compared with published numerical [46] and experimental [47] results in Fig. 6b. The comparisons are in good agreement and therefore, verify our calculations. Comparisons of mean Nusselt numbers calculated at different *Re* are shown in Table 2 and the maximum error in our calculations is around 3% as compared to the published results.

### 2.1.5 Convective heat transfer with moving immersed boundary

In order to validate the FSI solver for convective heat transfer with moving boundary, we consider a transversely oscillating heated cylinder of diameter $D = 1$ in a $40D \times 10D$ channel [48]. The cylinder is located midway between the channel walls at a distance of $10D$ from the inlet. The dimensionless oscillation frequency, amplitude of the transverse velocity of the cylinder, Reynolds number and Prandtl number are 0.2, 0.5, 200 and 0.71, respectively (case 2 in Ref. [48]). We employed a $256 \times 256$ non-uniform grid with $\Delta x_{min} = 0.01$ and $\Delta y_{min} = 0.01$, and $\Delta t = 0.0025$. The local Nusselt numbers at surface of the cylinder calculated in present work and by Fu and Tong [48] at different time instances are compared in Fig. 7. The present calculations are in good agreement with Fu and Tong [48] and thus verify the present numerical model for convective heat transfer with moving immersed boundary.

## 3  Results and Discussion

We demonstrate heat transfer enhancement via large-scale flow-induced deformation in the FSI benchmark proposed by Turek and Hron [8]. The flow boundary conditions and relevant parameters of the benchmark are discussed in section 2.1.1. The thermal boundary conditions and associated simulation parameters are described as follows. At the inlet and channel walls, Dirichlet boundary condition for temperature, $T_0 = 0$ and $T_w = 1$, were applied, respectively. Zero Neumann boundary condition was applied for temperature at the channel outlet and the fluid-structure interface. The temperature is zero initially in a $4.1D \times 41D$ channel and Prandtl number is taken as 1 in the simulation.

We utilize a $384 \times 160$ non-uniform grid with $\Delta x_{min} = 0.0231$ and $\Delta y_{min} = 0.02$, and $\Delta t = 0.01$. The grid convergence study was performed in Ref. [11] for the FSI benchmark and $256 \times 128$ grid was deemed adequate for resolving the flow field coupled with structure dynamics. In the present work, in order to resolve thermal field in the downstream of the channel, we perform the grid convergence study for three grids keeping the resolution near the cylinder and plate same as those in Ref. [11] and adding more grid points in the downstream. Four different grid sizes: $256 \times 128$ (baseline), $320 \times 160$ (fine), $384 \times 160$ (finer) and $512 \times 160$ (finest) are selected and differences in the amplitude (frequency) of the plate obtained using the fine grid with respect to baseline grid, finer grid with respect to fine grid and the finest grid with respect to the finer grid are 2.2% (2.6%), 0.0% (0.0%) and 0.0% (0.0%), respectively. Similarly, these differences in the mean Nusselt number, (eq. (11)), are 4.8%, 4.3% and 2%, respectively. Since deviations in the results obtained using the finer grid are small, we employ the finer grid ($384 \times 160$) in all simulations presented in this work.

## 3.1 Enhanced mixing of the fluid due to FSI

The vorticity contours are plotted in Fig. 8 after the flow field reaches time-periodic state (see also associated computer animation provided as supplementary data [49]). The four plots a-d in Fig. 8 correspond to four time instances in a typical cycle of the oscillation of the plate, as shown by the black dots in inset of Fig. 8. As shown in Fig. 8a-b, the fluid accelerates near the cylinder in the lower half of the channel due to the partial blockage by the plate in the upper half. The clockwise (Fig. 8b) and counter-clockwise (Fig. 8d) vortices generated due to the motion of the plate interact with the bottom and top channel wall, respectively. The flow induces a wave-like deformation in the plate and the plate attains self-sustained periodic oscillation with plateau amplitude after a short time [11]. The vorticity generated on the cylinder surface is almost annihilated by the wall vorticity from the wall closest to it, as evidenced by the residual vorticity footprint downstream. The cross-annihilation of vorticity of a particular sign with the vorticity of opposite sign generated on the cylinder or wall surfaces, results in descending strength of the vortices as they advect along the channel length. Thus, the wake vortices sweep the higher sources of vorticity generated on the channel walls out into the high velocity regions which aids in the mixing of the fluid in the bulk as well as near the channel walls.

### 3.2 Thermal augmentation due to FSI

The isotherms and instantaneous Nusselt numbers at channel walls are plotted in Fig. 9 and Fig. 10, respectively, after the flow as well as temperature field reach the time-periodic state (see also associated computer animation provided as supplementary data [49]). The four plots a-d in Fig. 9 and Fig. 10 corresponds to four time instances of the oscillation (black dots in inset of Fig. 8). The isotherms in Fig. 9 qualitatively indicate reduction in thermal boundary layer thickness at the locations of the convected vortices described in section 3.1. In order to confirm this trend quantitatively, we plot temperature profile at the four instances at $x = 8$ and 27 in Fig. 11. The change in temperature gradient at the walls at these instances confirms the effect of the convected vortices. The enhanced mixing of the fluid described in section 3.1 assists in augmenting convection in the bulk as well as near the channel walls. The peaks of instantaneous Nusselt numbers at channel walls ($x > 8$) in Fig. 10 are signature of reduction in thermal boundary layer thickness due to convected vortices. The height of the local peaks decrease due to decrease in strength of the vortices along the channel length as described in section 3.1. The influence of decreasing strength of vortices can be seen through temperature profiles plotted at $x = 8$ and 27 in Fig. 11a and b, respectively. The temperature gradients at the walls are larger at $x = 8$ as compared to those at $x = 27$, which implies larger Nusselt numbers at the former location. Overall, the vortices increase mixing in the channel as hot fluid near the channel walls moves towards the center of the channel and get replenished by relatively colder fluid. The interaction of the vortices with the channel wall helps in reducing thermal boundary layer.

The local fluid acceleration caused by the presence of cylinder also aids in improving heat transfer, however, this effect is regional. Due to upward motion of the plate (Fig. 8a-b), the fluid rushes through the lower half of the channel resulting in the fluid acceleration near the cylinder region, which reduces thermal boundary layer thickness (Fig. 9b, at $x \sim 6$) and improves Nusselt number ($Nu_{bw}$) regionally at bottom wall (Fig. 10b, at $x \sim 6$). Similarly, downward motion of the plate (Fig. 8b-d) improves Nusselt number ($Nu_{tw}$) regionally at top wall (Fig. 10d, at $x \sim 6$). The peaks at $x \sim 2$ for the top as well as bottom wall are caused by the accelerating fluid due to blockage by the cylinder (Fig. 10a-d).

## 3.3 Quantification of augmentation in heat transfer

In order to quantify the heat transfer enhancement achieved in the case described in section 3.2 - the flow past rigid cylinder attached with deformable plate in a heated channel (referred as CDP hereafter) -we performed simulations for three additional configurations listed in Table 1. These cases correspond to a channel flow without bluff body (CHL), channel flow past a rigid cylinder (CYL) and channel flow past a rigid cylinder attached with rigid plate (CRP), with same simulations parameters as those used in CDP. The vorticity contours and isotherms for all configurations at $t = 80$ are plotted in Fig. 12 and Fig. 13, respectively. We note that the vortices are present in CYL and CDP (Fig. 12b and d, respectively) and strength of the vortices decreases in stream-wise direction. However the vortices are stronger, closer to the walls and convect much further in the downstream in CDP as compared to those in CYL. The flow is steady in CHL and the rigid plate inhibits flow-instabilities to suppress the vortex shedding in CRP. As illustrated by isotherms in CYL and CDP in Fig. 13b and d, respectively, interaction of vortices with channel wall helps in reducing thermal boundary layer and thereby augmenting heat transfer. In order to quantify the augmentation, we plot stream-wise variation of time-averaged Nusselt number ($Nu_{avg}$, eq. (10)) for the above cases in Fig. 14. We note that the $Nu_{avg}$ decreases monotonically along the channel for CHL case while a local peak occurs at the location of the cylinder ($x \sim 2$) in rest of other cases. The local peaks occur due to partial blockage of the flow by the cylinder, as explained in section 3.2. The vortex shedding in CYL helps in improving the Nusselt number at all stream-wise locations and $Nu_{avg}$ improves by 18% at $x \sim 40$, as compared to that in CHL and CRP. The results indicate significant increase in $Nu_{avg}$ in CDP due to enhanced convection as explained in sections 3.1 and 3.2. With respect to CHL, the maximum enhancement is around 137% at $x \sim 10.6$ and 57% at channel outlet ($x \sim 40$). The regional thermal improvement, for instance at $x \sim 10.6$, implies that large-scale deformation could be leveraged to cool spatially-varying peak thermal loading on a surface, for instance, hotspots on electronic chips [50]. The time- and space-averaged Nusselt number ($Nu_{mean}$, eq. (11)) for all four configurations is compared using bar charts in

Fig. 15a (first set of red bars) and Nusselt number is 69%, 41% and 54% larger in CDP as compared to that in CHL, CYL and CRP, respectively. We note that the augmentation in CDP is the largest; however, power required in pumping the fluid may be different in these cases. We define dimensionless pumping power ($P$) required to maintain a given flow rate $Q$ in a channel of width $H$, as follows,

$$P = Q\left[\int_0^H p_{inlet}(y)dy - \int_0^H p_{outlet}(y)dy\right] \quad (13)$$

where $p_{inlet}$ and $p_{outlet}$ are pressures per unit span wise length across the channel inlet and outlet, respectively. The flow rate $Q$ per unit span wise length is same in all cases and defined as,

$$Q = \int_0^H u(y)dy \quad (14)$$

In time-periodic flows, the average pumping power ($P_{avg}$) is time-average of the instantaneous pumping power over a period of oscillation. The dimensionless pumping power required in different cases is plotted in

Fig. 15b (first set of red bars), which shows that the thermal enhancements in CYL, CRP and CDP are achieved at expense of more energy required in pumping of the fluid in the channel. It is worthwhile to compare the effectiveness of CYL, CRP and CDP with respect to CHL and we use the following metrics for the comparison among these configurations. We define efficiency index ($\eta$) [51],

$$\eta = \frac{\eta_h}{\eta_f} \quad (15)$$

where $\eta_h$ and $\eta_f$ are enhancement factors for the heat transfer and pumping power, respectively, defined as follows [51],

$$\eta_h = \frac{Nu_{mean}}{Nu_{mean} \text{ for channel}} \quad (16)$$

$$\eta_f = \frac{P_{avg}}{P_{avg} \text{ for channel}} \quad (17)$$

We plot the efficiency indices for CYL, CRP and CDP in

Fig. 15c (first set of red bars), which shows that CDP is 10% and 6% more efficient than CYL and CRP, respectively. Therefore, large-scale flow-induced deformation helps in the thermal augmentation as compared to a case in which rigid bluff body is employed as vortex generator.

### 3.3.1 Effect of Prandtl number

In this and the following sections, we perform additional simulations to assess the influence on Prandtl number ($Pr$), Reynolds number ($Re$), Young's Modulus ($E$) and channel length. In order to assess the effect of $Pr$, we increase it to $Pr = 2$, in CHL, CYL, CDP and CYL keeping all other parameters same. $Nu_{mean}$, $P_{avg}$ and $\eta$ for these four simulations are plotted using second set of green bars in

Fig. 15a, b, and c, respectively. At $Pr = 2$, the $Nu_{mean}$ increases in all configurations as compared to the respective baseline cases (first set of red bars vs. second set of respective green bars in

Fig. 15a) due to decrease in thermal boundary layer thickness at larger Prandtl number [39]. However, pumping power is same as compared to the respective baseline cases and the improvement in $\eta$ in CDP ($Pr = 2$) is 6% as compared to CYL ($Pr = 2$), which is lower than that in the baseline case (10%, first set of red bars in

Fig. 15a). Therefore, larger Prandtl number assists in achieving the enhancement, however, it is less efficient than the baseline case (CDP, $Pr = 1$).

### 3.3.2 Effect of Reynolds number

The effect of Reynolds number is studied by increasing it to $Re = 200$ and 500 in CHL, CYL, CDP and CYL keeping all other parameters same. $Nu_{mean}$, $P_{avg}$ and $\eta$ for these eight simulations are plotted in

Fig. 15a, b, and c, respectively. Third set of blue bars and fourth set of magenta bar show results for $Re = 200$ and 500, respectively. At $Re = 200$, $Nu_{mean}$ improves due to larger convection, pumping power reduces due to smaller pressure drop needed at larger $Re$, however, the efficiency index decreases for all configurations (compare first set of red bars vs. third set of respective blue bars in

Fig. 15a, b, and c, respectively). Note that $\eta$ decreases because relative to CHL case the increase in $Nu_{mean}$ for other cases is not in proportion to the decrease in $P_{avg}$. In particular, vortex shedding (not shown in the paper) occurs behind the tip of the rigid plate in CRP case as compared to that in baseline CRP ($Re = 100$). The efficiency index in the CDP is around 23% more compared to CYL ($\eta = 0.30$ to $0.37$) for $Re = 200$, which is larger than the baseline case (10%, $\eta = 0.45$ to $0.50$). We see similar trends for Re = 500 with forth set of magenta bars in

Fig. 15. Larger thermal enhancement is achieved with lower pumping power at $Re = 500$ as compared to $Re = 200$. $Nu_{mean}$ increases by around 59% and $\eta$ deceases by 40% for the former as compared to the latter in the CDP case. Overall, larger Reynolds number helps in achieving larger thermal augmentation at lower pumping power, however, it is less efficient than the baseline case (CDP, $Re = 100$).

### 3.3.3 Effect of Young's Modulus

We increase Young's Modulus to $E = 2.8 \times 10^3$ in CDP and keep all other parameters same in CHL, CYL, CRP and CDL. $Nu_{mean}$, $P_{avg}$ and $\eta$ for these four simulations are plotted using fifth set of black bars in

Fig. 15a, b, and c, respectively. Since Young's Modulus is associated with the deformation of the plate, the results in only CDP are influenced. Due to 17% reduction in the amplitude of the oscillating plate [11], $Nu_{mean}$ and $P_{avg}$ decrease, which results in lower $\eta$. $\eta$ in the CDP is same as that in CYL at $E = 2.8 \times 10^3$. Hence, larger Young's Modulus impedes the enhancement as well as costlier with respect to CHL. Overall, the Nusselt number ($Nu_{mean} = 42.7$) and efficiency index ($\eta = 0.5$) are the largest for CDP with $Re = 500$, $Pr = 1$, $E = 1.4 \times 10^3$ and CDP with $Re = 100$, $Pr = 1$, $E = 1.4 \times 10^3$, respectively for the simulations presented in

Fig. 15.

### 3.3.4 Effect of channel length

Finally, we examined effect of channel length by varying it to $L = 61$ in the baseline case (CDP, $L = 41$, $Re = 100$, $Pr = 1$, $E = 1.4 \times 10^3$, $Nu_{mean} = 18.71$). The mean Nusselt number decreases by 5.5% ($Nu_{mean} = 16.69$) for 49% longer channel ($L = 61$). Since strength of the vortices decrease as they convect in the downstream, the mean Nusselt number decreases marginally along the channel length. The pumping power increases by 47% in the latter case and the efficiency index is 0.49, same as in the baseline case. Note that the Nusselt number is dependent on the channel length and plateau Nusselt number for an infinitely long channel is 7.54 [39]. Thus, a finite channel length should be employed in order to enhance the heat transfer.

## 4　Conclusions

We demonstrate effective thermal enhancement in a laminar, incompressible heated channel flow via large-scale flow-induced deformation of an elastic thin plate attached on lee side of a rigid cylinder. The in-house fluid-structure interaction (FSI) solver with convective heat transfer is based on a sharp-interface immersed boundary method. The flow solver is strongly-coupled with an open-source structure dynamics solver using partitioned approach. In the present work, several validations are carried out against benchmark examples of conduction and convective heat transfer including moving structure boundaries, in order to test the convective heat transfer module of the FSI solver. The mechanism to achieve thermal augmentation via large-scale flow-induced deformation is as follows. The wake vortices generated due to the self-sustained motion of the plate sweep higher sources of vorticity generated on the channel walls out into the high velocity regions which aids in the mixing of the fluid in the bulk as well as near the channel walls. Due to the enhanced mixing, the hot fluid near the channel walls moves towards the center of the channel and get replenished by relatively colder fluid. The interaction of the vortices with the channel wall helps in reducing thermal boundary layer and thereby increasing Nusselt number at the channel walls. The improvement is quantified via testing of three additional configurations, namely, channel flow without bluff body; channel flow with a stationary and rigid cylinder, and channel flow with a stationary and rigid cylinder with a rigid plate attached on its lee side. Simulation results suggest that larger Prandtl number and Reynolds number promote the thermal enhancement, however, with less efficiency. On the other hand, larger Young's Modulus impedes the enhancement as well as requires larger pumping power. The numerical data presented will help to design better thermal augmentation systems involving large-scale flow-induced deformation of thin structures. Overall, the large-scale flow-induced deformation harnesses available flow energy and does not require any external power sources to achieve the thermal augmentation**.** The large regional thermal improvements could be leveraged to cool non-uniform peak thermal loading on a surface.

## 5　Acknowledgements

R.B. gratefully acknowledges financial support from Department of Science and Technology (DST), New Delhi through fast track scheme for young scientists and thanks Prof. Rajat Mittal at Johns Hopkins University for useful discussions. This work was also partially supported by an internal grant from Industrial Research and Consultancy Centre (IRCC), IIT Bombay.

## 6 Nomenclature

| | |
|---|---|
| $D$ | Diameter of the cylinder |
| $H$ | Height of the channel |
| $E$ | Young's modulus of the plate ($E^*/(\rho_f^* U_m^{*2})$) |
| $L$ | Channel length |
| $P_{avg}$ | Average pumping power ($P_{avg}^*/(\rho_f^* D^* U_m^{*3})$) |
| $p$ | Pressure ($p^*/(\rho_f^* U_m^{*2})$) |
| $T$ | Temperature (eq. 4) |
| $U_m$ | Mean velocity at the left boundary of the channel |
| $v_i$ | Component of dimensionless velocity vector |
| $Y_{tip}$ | Maximum Y-displacement of the tip of the plate ($Y_{tip}^*/D^*$) |

*Greek symbols*

| | |
|---|---|
| $\rho$ | Density [kg m$^{-3}$] |
| $\mu$ | Dynamic viscosity [N m$^{-2}$ s] |
| $\nu$ | Kinematic viscosity [m$^2$/s] |
| $\alpha$ | Thermal diffusivity [N m$^{-2}$ s] |
| $\omega_z$ | Vorticity component in $z$ direction |
| $\eta$ | Efficiency index of heat transfer enhancement |
| $\eta_h$ | Heat transfer enhancement factor with respect to channel flow without bluff body |
| $\eta_f$ | Pumping power with respect to channel flow without bluff body |
| $\Delta t$ | Dimensionless time-step |

*Subscripts*

| | |
|---|---|
| 0 | Initial/inlet |
| w | Wall |
| bw | Bottom wall |

*tw*   Top wall

*avg*   Time-average value

mean   Time- and space-average value

*f*   Fluid

*s*   Structure

*Superscripts*

\*   Dimensional quantity

*Dimensionless numbers*

*Re*   Reynolds number ($\rho_f^* U_m^* D^* / \mu$)

*Pr*   Prandtl number ($\nu/\alpha$)

*Pe*   Peclet number (*RePr*)

*Nu*   Nusselt number (eq. 5)

*Acronyms*

1D   One-dimensional

2D   Two-dimensional

3D   Three-dimensional

FSI   Fluid-Structure interaction

# 7   References

[1] Beeby S.P., Tudor M. J. and White N. M., Energy harvesting vibration sources for microsystems applications, Meas. Sci. Technol. Vol. 17,175–195 (2006).

[2] Weissenbock, N.M., C.M. Weiss, H.M. Schwammer, and H. Kratochvil, Thermal windows on the body surface of African elephants (Loxodonta africana) studied by infrared thermography. Journal of Thermal Biology, 2010. **35**(4): p. 182-188.

[3] Mills, Z.G., B. Aziz, and A. Alexeev, Beating synthetic cilia enhance heat transport in microfluidic channels. Soft Matter, 2012. **8**(45): p. 11508-11513.

[4] Shelley M. J. and Jun Zhang, Flapping and Bending Bodies Interacting with Fluid Flows, Annual Review of Fluid Mechanics, Vol. 43: 449-465, 2011.

[5] Zheng, X., Q. Xue, R. Mittal, and S. Beilamowicz, A Coupled Sharp-Interface Immersed Boundary-Finite-Element Method for Flow-Structure Interaction With Application to Human Phonation. Journal of Biomechanical Engineering-Transactions of the ASME, 2010. 132(11).

[6] Baiijens, F. P. T., "A fictitious domain mortar element method for fluid–structure interaction" *Int. J. Numer. Meth. Fluids*, vol 35, pp. 743–761, 2001.

[7] Vigmostad S. C., Udaykumar H. S., Lu J. and Chandran K. S., "Fluid–structure interaction methods in biological flows with special emphasis on heart valve dynamics," Int. J. Numer. Meth.Biomed.Engng. vol. 26, pp. 435-470, 2010.

[8] Turek, S. and J. Hron, Proposal for Numerical Benchmarking of Fluid-Structure Interaction between an Elastic Object and a Laminar Incompressible Flow. *Lecture Notes in Comp Sc & Eng. (Eds: Bungartz H.-J. & Schaefer M.)*, Springer Verlag. 2006

[9] Dunne T. and Rannacher R. (2006). Adaptive Finite Element Approximation of fluid-structure interaction based on an Eulerian Variational Formulation. *Lecture Notes in Comp Sc & Eng. (Eds: Bungartz H.-J. & Schaefer M.)*, Springer Verlag.

[10] Heil M., Hazel A. L. and Boyle J. (2008). Solvers for large-displacement fluid–structure interaction problems: segregated versus monolithic approaches. Comput. Mech. Vol. 43, 91-101.

[11] Bhardwaj, R. and R. Mittal, Benchmarking a coupled Immersed-Boundary-Finite-Element solver for large-scale flow-induced deformation. AIAA, 2012. **50**: p. 1638-1642.

[12] Lee J. and You D. Study of vortex-shedding-induced vibration of a flexible splitter plate behind a cylinder. Phys. Fluids 25, 110811 (2013).

[13] Tian, F. B., Dai, H., Luo, H., Doyle, J. F., Rousseau, B., "Fluid-Structure interaction involving large deformations: 3D simulations and applications to biological systems," J. of Comp. Phy., vol. 258, pp. 451-569, 2014.

[14] Webb RL and Kim N-H, Principles of enhanced heat transfer. 2005: Taylor & Francis, New York.

[15] Bergles, A.E., Recent developments in enhanced heat transfer. Heat and Mass Transfer, 2011. **47**(8): p. 1001-1008.

[16] Valencia, A., Heat transfer enhancement in a channel with a built-in rectangular cylinder. Heat and Mass Trasnfer, 1995. **30**: p. 423-427.

[17] Yoon, D.H., K.S. Yang, and C.B. Choi, Heat Transfer Enhancement in Channel Flow Using an Inclined Square Cylinder. Journal of Heat Transfer-Transactions of the ASME, 2009. **131**(7).

[18] Chatterjee, D. and B. Mondal, Forced convection heat transfer from an equilateral triangular cylinder at low Reynolds numbers. Heat and Mass Transfer, 2012. **48**(9): p. 1575-1587.

[19] Biswas, G. and H. Chattopadhyay, Heat-transfer in a channel with built-in wing-type vortex generators. International Journal of Heat and Mass Transfer, 1992. **35**(4): p. 803-814.

[20] Hiravennavar, S.R., E.G. Tulapurkara, and G. Biswas, A note on the flow and heat transfer enhancement in a channel with built-in winglet pair. International Journal of Heat and Fluid Flow, 2007. **28**(2): p. 299-305.

[21] Fiebig, M., U. Brockmeier, N.K. Mitra, and T. Guntermann, Structure of velocity and temperature-fields in laminar channel flows with longitudinal vortex generators. Numerical Heat Transfer, 1989. **15**(3): p. 281-302.
19

[22] A.W. Fan, J.J. Deng, J. Guo, W. Liu, A numerical study on thermo-hydraulic characteristics of turbulent flow in a circular tube fitted with conical strip inserts, Applied Thermal Engineering, 31: 2819-2828, 2011.

[23] A.W. Fan, J.J. Deng, A. Nakayama, W. Liu, Parametric study on turbulent heat transfer and flow characteristics in a circular tube fitted with louvered strip inserts, International Journal of Heat and Mass Transfer, 55: 5205-5213, 2012.

[24] Wang, C.C. and C.K. Chen, Forced convection in a wavy-wall channel. International Journal of Heat and Mass Transfer, 2002. **45**(12): p. 2587-2595.

[25] Celik, B., M. Raisee, and A. Beskok, Heat transfer enhancement in a slot channel via a transversely oscillating adiabatic circular cylinder. International Journal of Heat and Mass Transfer, 2010. **53**(4): p. 626-634.

[26] Beskok, A., M. Raisee, B. Celik, B. Yagiz, and M. Cheraghi, Heat transfer enhancement in a straight channel via a rotationally oscillating adiabatic cylinder. International Journal of Thermal Sciences, 2012. **58**: p. 61-69.

[27] Khanafer, K., A. Alamiri, and I. Pop, Fluid-structure interaction analysis of flow and heat transfer characteristics around a flexible microcantilever in a fluidic cell. International Journal of Heat and Mass Transfer, 2010. **53**(9-10): p. 1646-1653.

[28] Soti AK and Bhardwaj R, Numerical study of heat transfer enhancement via flow-induced deformation of elastic plate in channel flow, Proceedings of the 22nd National and 11th International ISHMT-ASME Heat and Mass Transfer Conference, December 28-31, 2013, IIT Kharagpur, India.

[29] Shi J, Hu J, Schafer S R, Chen C.L. Numerical study of heat transfer enhancement of channel via vortex-induced vibration, Applied Thermal Engineering, pp 838–845, 2014.

[30] Mittal, R., Iaccarino, G. Immersed Boundary Methods. *Ann Rev Fluid Mech*, 2005, Vol 37, pp 239-61.

[31] Mittal, R., H. Dong, M. Bozkurttas, F.M. Najjar, A. Vargas, and A. von Loebbecke, A versatile sharp interface immersed boundary method for incompressible flows with complex boundaries. Journal of Computational Physics, 2008. **227**(10): p. 4825-4852.

[32] Seo, J.H. and R. Mittal, A sharp-interface immersed boundary method with improved mass conservation and reduced spurious pressure oscillations. Journal of Computational Physics, 2011. 230(19): p. 7347-7363.

[33] Mittal, R., X. Zheng, R. Bhardwaj, J.H. Seo, Q. Xue, and S. Bielamowicz, Toward a simulation-based tool for the treatment of vocal fold paralysis. Frontiers in physiology, 2011. **2**.

[34] Zang, Y., Streett, R. L., and Koseff, J. R. (1994). A non-staggered fractional step method for time-dependent incompressible Navier-Stokes equations in curvilinear coordinates. J. Comput. Phys. 114(1), 18-33.

[35] Chorin, A. J. Numerical solution of the navier-stokes equations. Math. Comput. 22:745–762, 1968.

[36] W.H. Press, S.A. Teukolsky, W.T. Vetterling, and B.P. Flannery (1992), *Numerical Recipes: The Art of Scientific Computing*, Second Edition, (Cambridge University Press, New York, pp 818-880.

[37] Tahoe is an open source C++ finite element solver, which was developed at Sandia National Labs, CA (http://sourceforge.net/projects/tahoe/).

[38] Bhardwaj R, Zeigler K, JH Seo, and KT Ramesh, Nguyen TD, A Computational Model of Blast Loading on Human Eye, Biomechanics and Modeling in Mechanobiology, Vol. 13 (1), pp 123-140, 2014.

[39] Bejan, A., Convection Heat Transfer. 3rd edition. 2004, Wiley. Hoboken, N.J.

[40] Fung, Y. C. Foundations of Solid Mechanics, 2nd ed. 1993. Springer-Verlag, New York.

[41] Incropera, F.P., D.P. DeWitt, T.L. Bergman, and A.S. Lavine, Fundamentals of Heat and Mass Transfer. Vol. 7. 2011.

[42] MATLAB PDE Toolbox Release 2009a, The MathWorks, Inc., Natick, MA. http://www.mathworks.in/products/pde/

[43] Knudsen, J.D., Katz, D.L., 1958. Fluid Dynamics and Heat Transfer. McGraw Hill, NewYork.

[44] Zhuauskas, A., 1972. Heat transfer from tubes in cross-flow. In: Harnett, J.P., Irwine, T.F. (Eds.), Advances in Heat Transfer. Academic Press, New York.

[45] Churchill, S.W., Bernstein, M.J., 1977. A correlating equation for forced convection from gases and liquids to a circular cylinder in crossflow. J. Heat Transfer 99, 300–306.

[46] Patnana, V.K., R.P. Bharti, and R.P. Chhabra, Two-dimensional unsteady forced convection heat transfer in power-law fluids from a cylinder. International Journal of Heat and Mass Transfer, 2010. 53(19-20): p. 4152-4167.

[47] Eckert, E.R.G. and E. Soehngen, Distribution of heat transfer coefficients around circular cylinders in cross flow at Reynolds numbers from 20 to 500. Trans. ASME, 1952. 74: p. 343–347.

[48] Fu, W.S. and B.H. Tong, Numerical investigation of heat transfer from a heated oscillating cylinder in a cross flow. International Journal of Heat and Mass Transfer, 2002. 45(14): p. 3033-3043.

[49] Computer animation provided as supplementary data, filename= Soti_bhardwaj_Sheridan.avi.

[50] Hamann, H.F., Weger, A., Lacey, J.A., Hu Z., Bose P., Cohen, E.,Wakil, J.., Hotspot-Limited Microprocessors: Direct Temperature and Power Distribution Measurements, IEEE Journal of Solid-State circuits. 2007 Vol. 42 (1), pp. 56-64.

[51] Yang S.-J., Numerical study of heat transfer enhancement in a channel flow using an oscillating vortex generator. Heat and Mass Transfer 39 (2003) 257–265

# 8 Tables

Table 1. Cases considered by Soti and Bhardwaj [28] and in present work to quantify heat transfer augmentation via flow-induced deformation

| Cases | Configuration |
|---|---|
| CHL | Channel flow without bluff body. |
| CYL | Channel flow with a stationary and rigid cylinder. |
| CRP | Channel flow with a stationary and rigid cylinder with a rigid plate attached on its lee side. |
| CDP | Channel flow with a stationary and rigid cylinder with a deformable plate attached on its lee side. |

Table 2. Time- and space- averaged Nusselt number ($Nu_{mean}$) for flow around a stationary heated cylinder

| $Re$ | Present work | Knudsen and Katz [43] | Zhuauskas [44] | Churchill and Bernstein [45] |
|---|---|---|---|---|
| 80 | 4.58 | 4.67 | 4.56 | 4.64 |
| 100 | 5.14 | 5.19 | 5.10 | 5.16 |
| 120 | 5.66 | 5.65 | 5.59 | 5.62 |
| 200 | 7.39 | 7.16 | 7.21 | 7.19 |

## 9 Figures

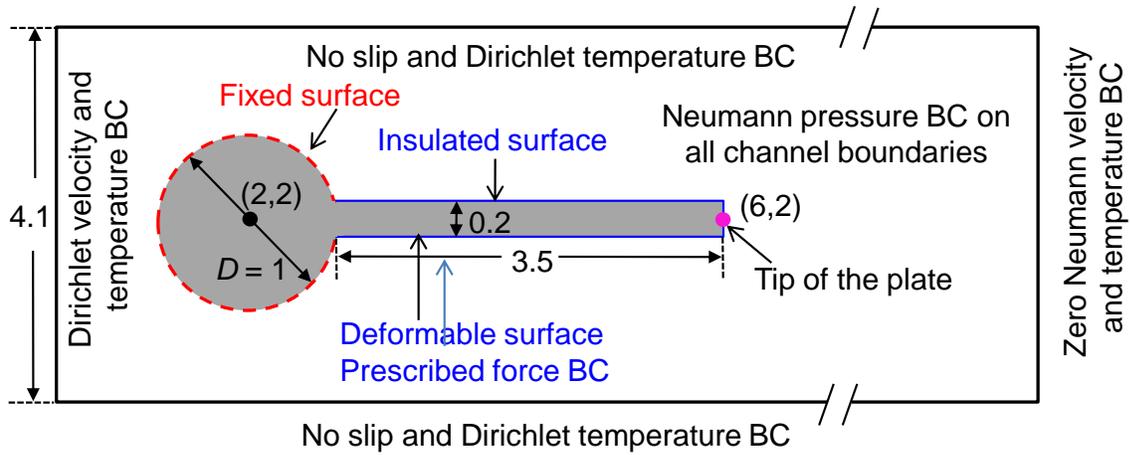

Fig. 1. Schematic and boundary conditions (BC) of the FSI benchmark problem with heat transfer. The FSI benchmark was first proposed by Turek and Hron [8] and later Bhardwaj and Mittal [11] validated their FSI solver with the benchmark. In present work, coupled convective heat transfer is considered with insulated cylinder and plate in a heated channel. Adapted with permission from [11]. Copyright (2011) Professor Rajat Mittal.

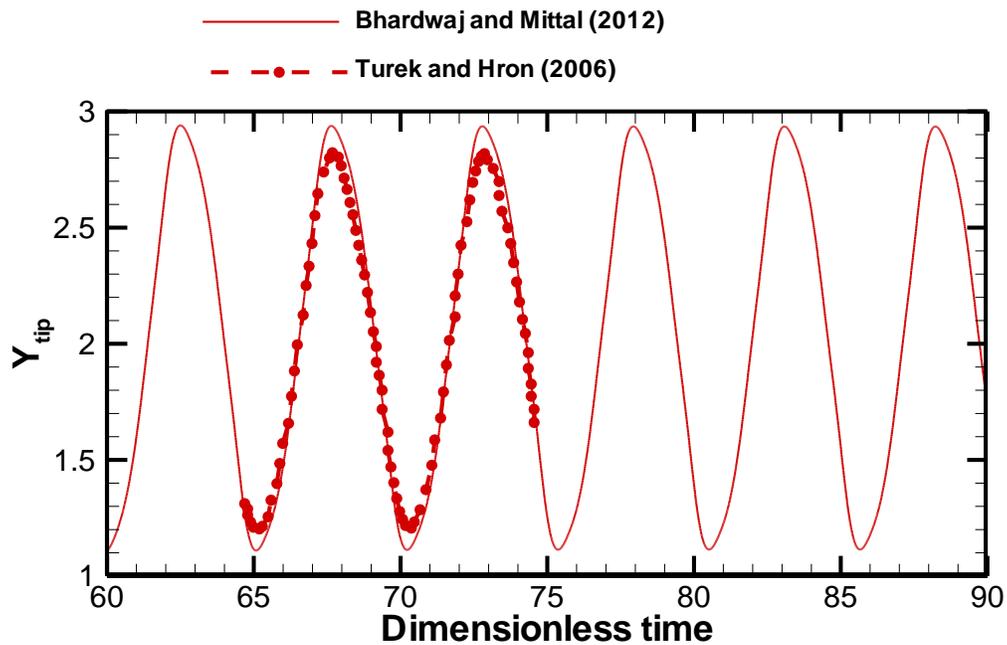

Fig. 2. Comparison between the present work and published results of Turek and Hron [8] for the stationary-state time-variation of X and Y displacement of the tip of the plate. Adapted with permission from [11]. Copyright (2011) Professor Rajat Mittal.

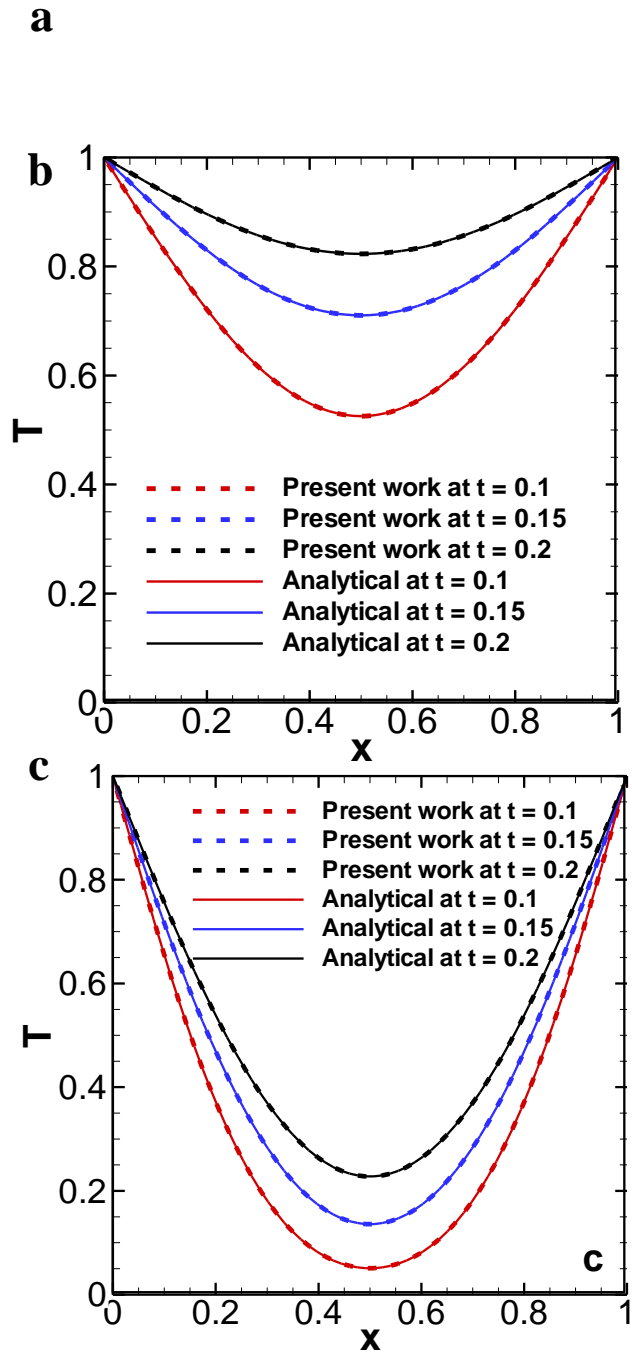

Fig. 3. Schematic of 1D conduction heat transfer in a slab with initial and boundary conditions (a). Comparisons of simulated time- and space-varying temperature in the slab for Peclet numbers, $Pe = 1$ (b), and $Pe = 4$ (c) with 1D analytical theory.

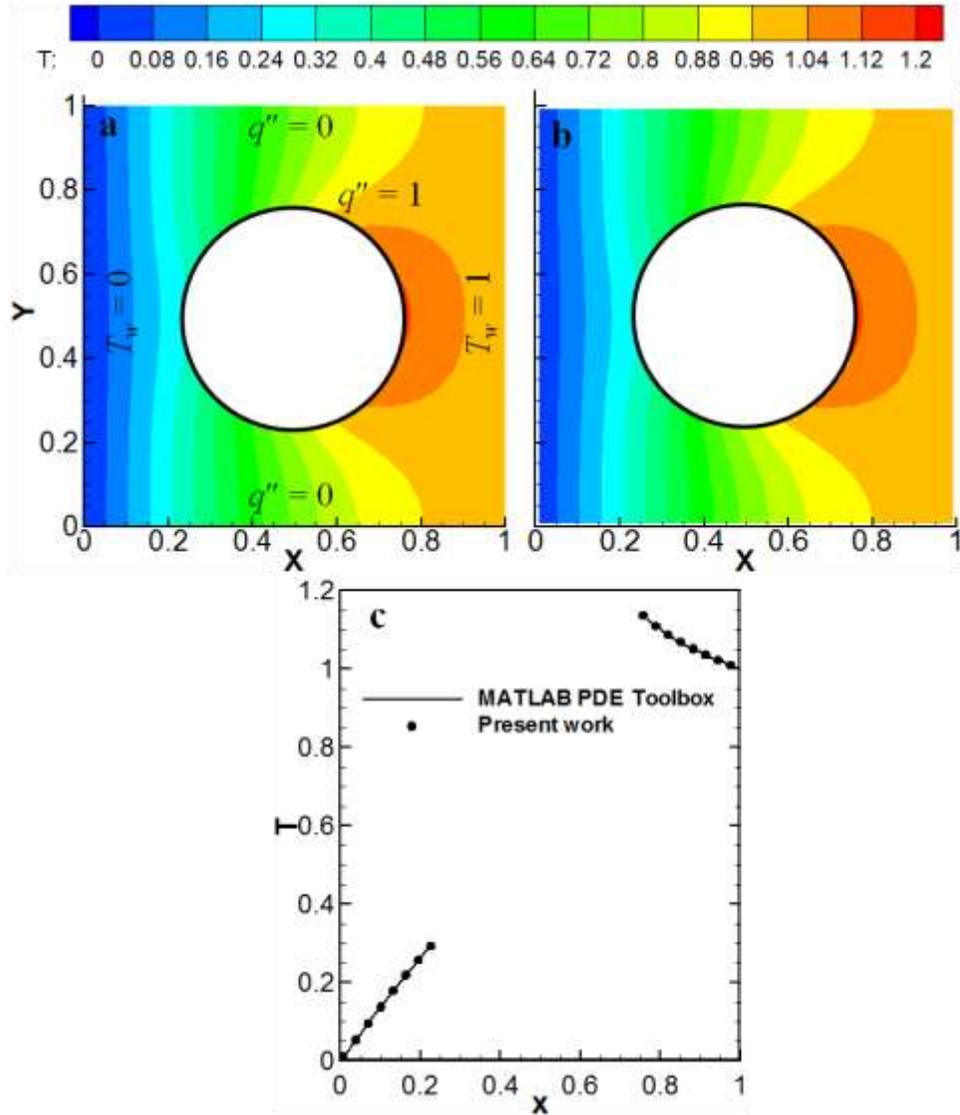

Fig. 4. 2D conduction heat transfer in a square block with boundary conditions shown in (a). Qualitative comparison of isotherms for *Pe* = 1 obtained using (a) MATLAB PDE Toolbox [42], and (b) the present numerical model. (c) Comparison of steady-state *x*-varying temperature at *y* = 0.5.

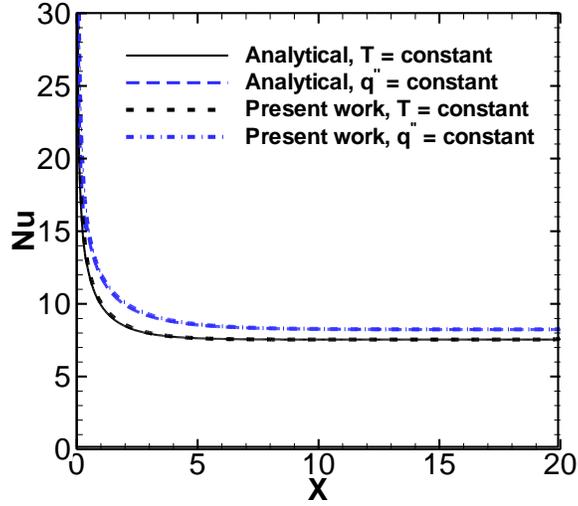

Fig. 5. Comparison of local Nusselt number along channel length with analytical theory. Two cases, constant wall temperature and constant wall flux are considered.

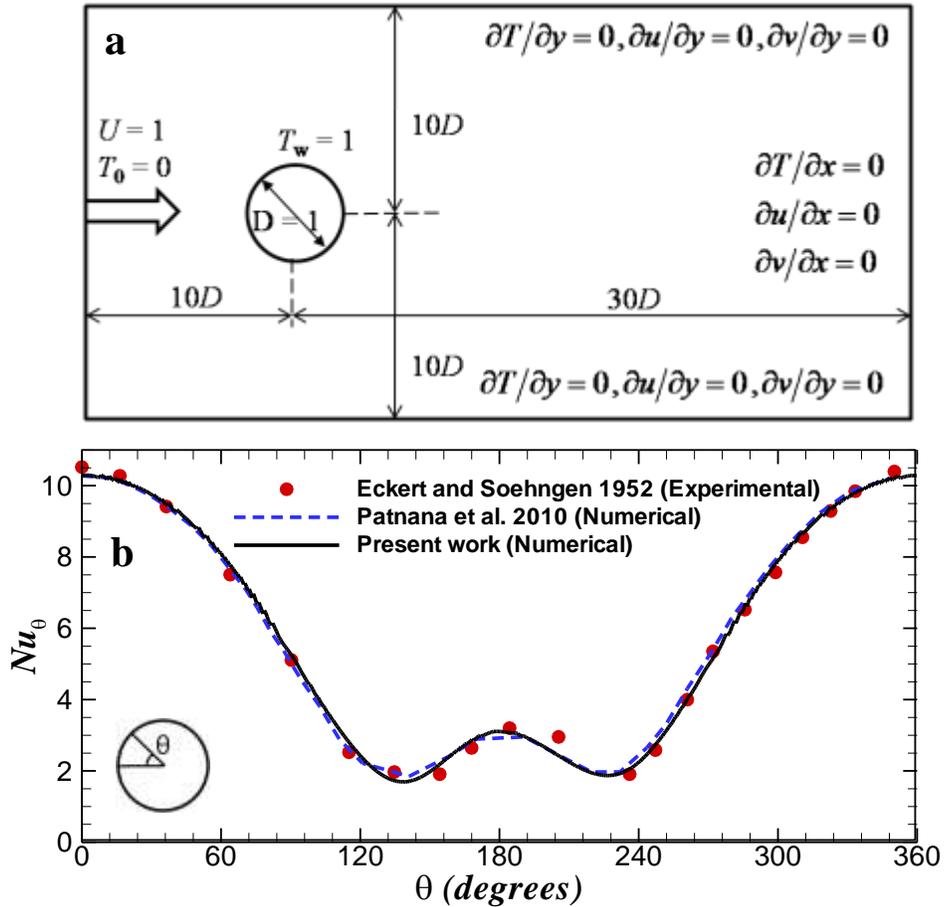

Fig. 6. a) Schematic and boundary conditions for flow past a heated and stationary cylinder. b) Comparison of local Nusselt number for $Re = 120$, $Pr = 0.7$ as function of azimuthal angle (shown in inset) at the surface of the cylinder with published numerical and experimental results.

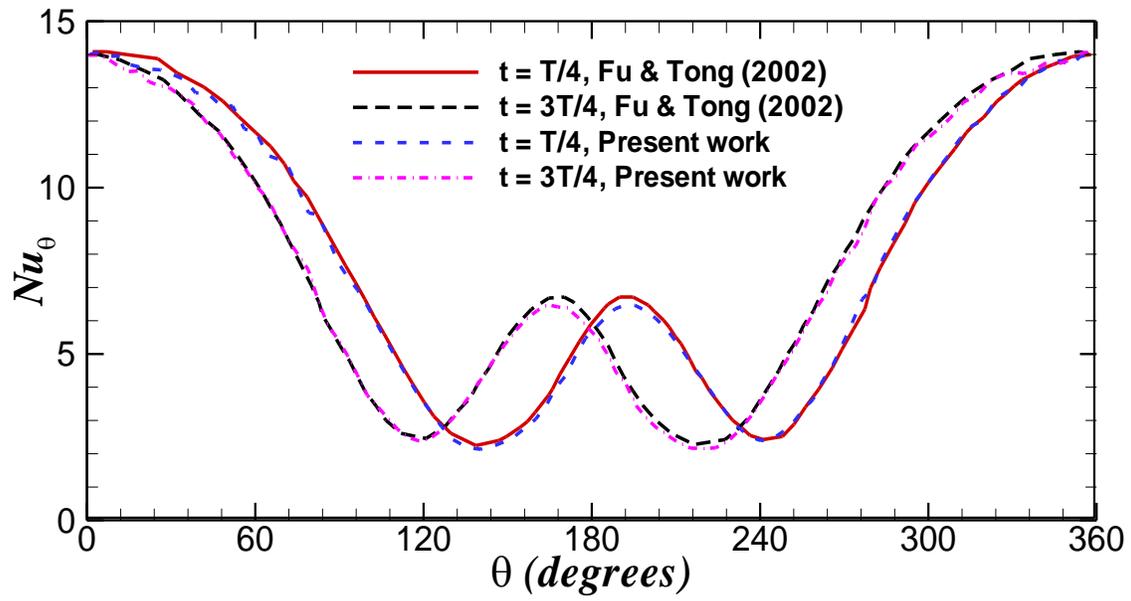

Fig. 7. Comparison of local Nusselt number as function of azimuthal angle (shown in inset of Fig 6b) at the surface of a transversely oscillating cylinder with numerical results of Fu and Tong [48] for $Re = 200$ and $Pr = 0.71$. Results are compared at different time instances, t = $T/4$ and $3T/4$, where $T$ is the time-period of the oscillation of the cylinder.

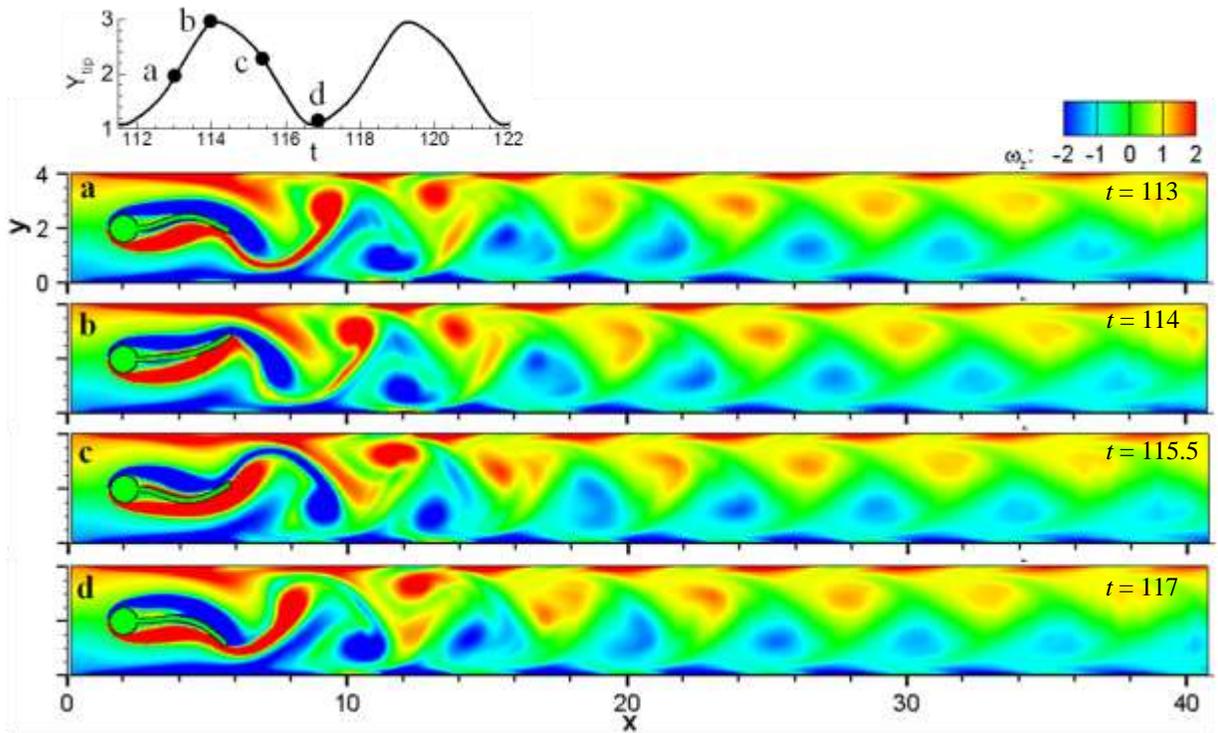

Fig. 8. Vorticity contours in a channel with a cylinder attached to an elastic plate at different time instances for $Re = 100$. The time instances are shown in the inset as black dots on time-varying position of the tip of the plate during a typical cycle of the plate oscillation.

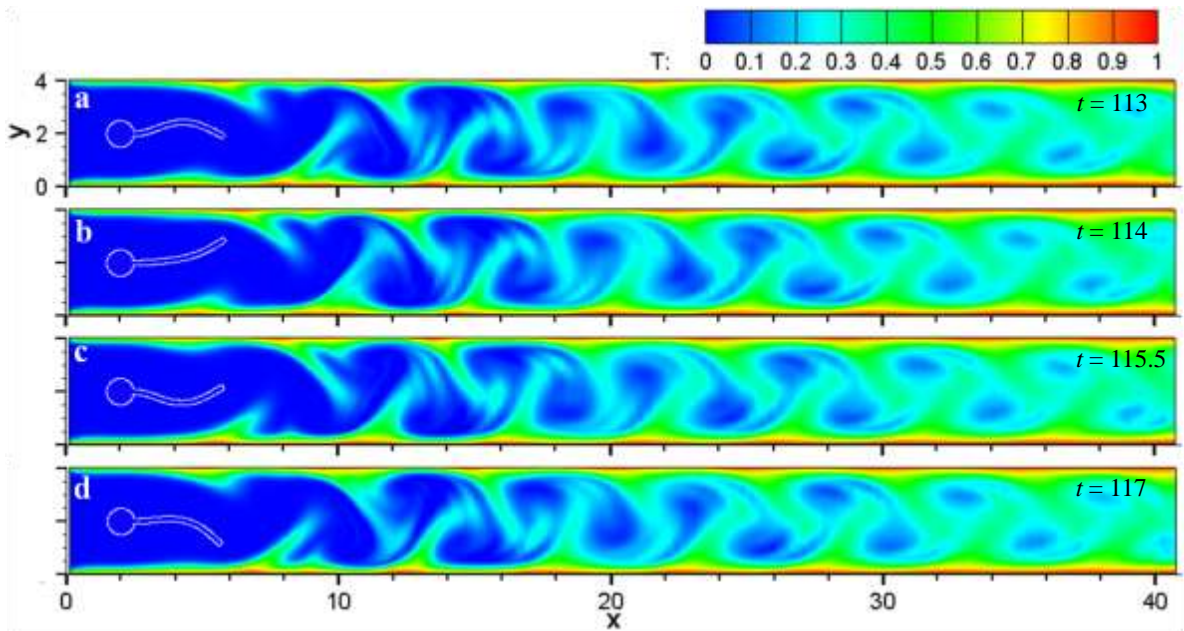

Fig. 9. Isotherms in a channel with a cylinder attached to an elastic plate at different time instances for $Re = 100$ and $Pr = 1$. The time instances are shown in the inset of Fig. 8 as black dots on time-varying position of the tip of the plate during a typical cycle of the plate oscillation.

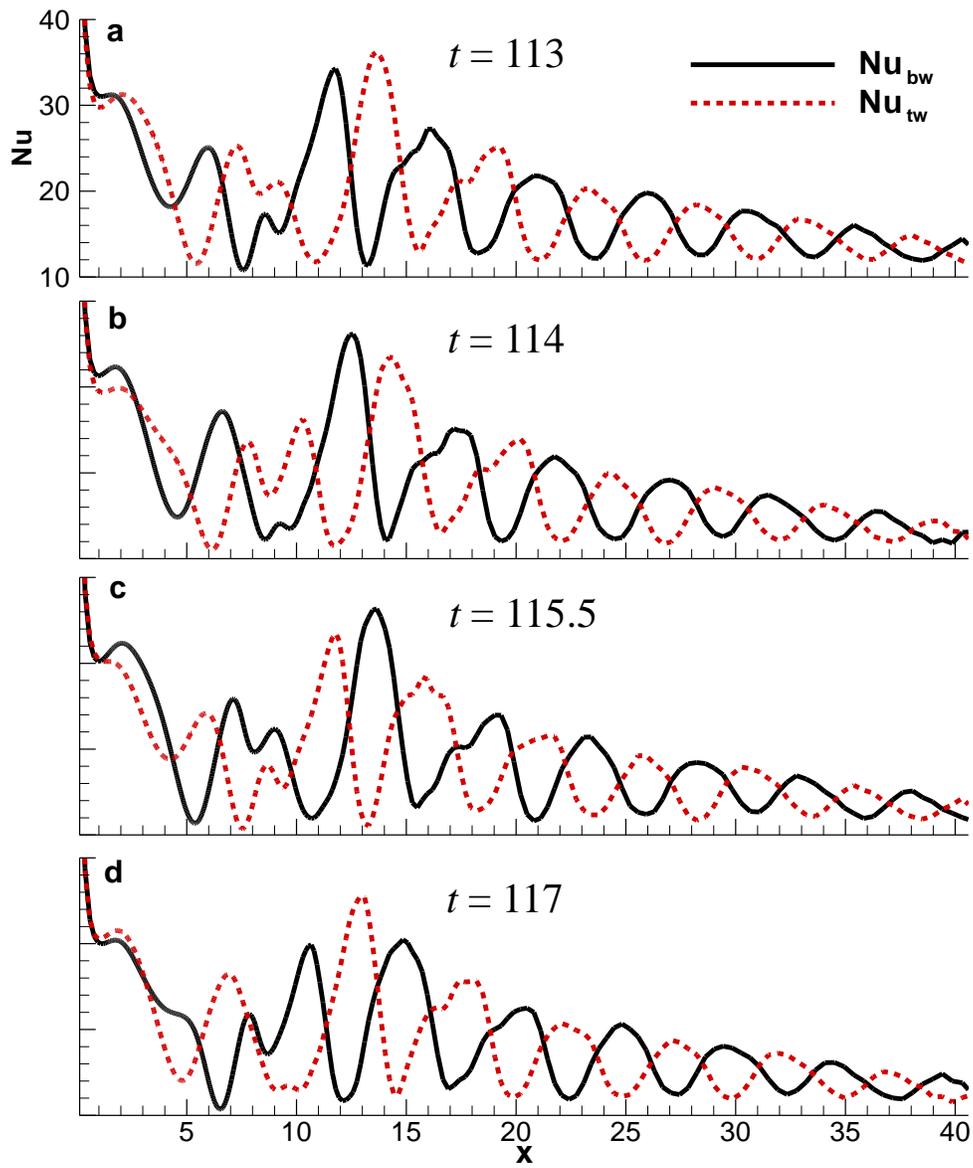

Fig. 10. Instantaneous Nusselt numbers at the bottom ($Nu_{bw}(x, t)$) and top ($Nu_{tw}(x, t)$) channel walls at different time instances, $Re = 100$, $Pr = 1$. The time instances are shown in the inset of Fig. 8 as black dots on time-varying position of the tip of the plate during a typical cycle of the plate oscillation.

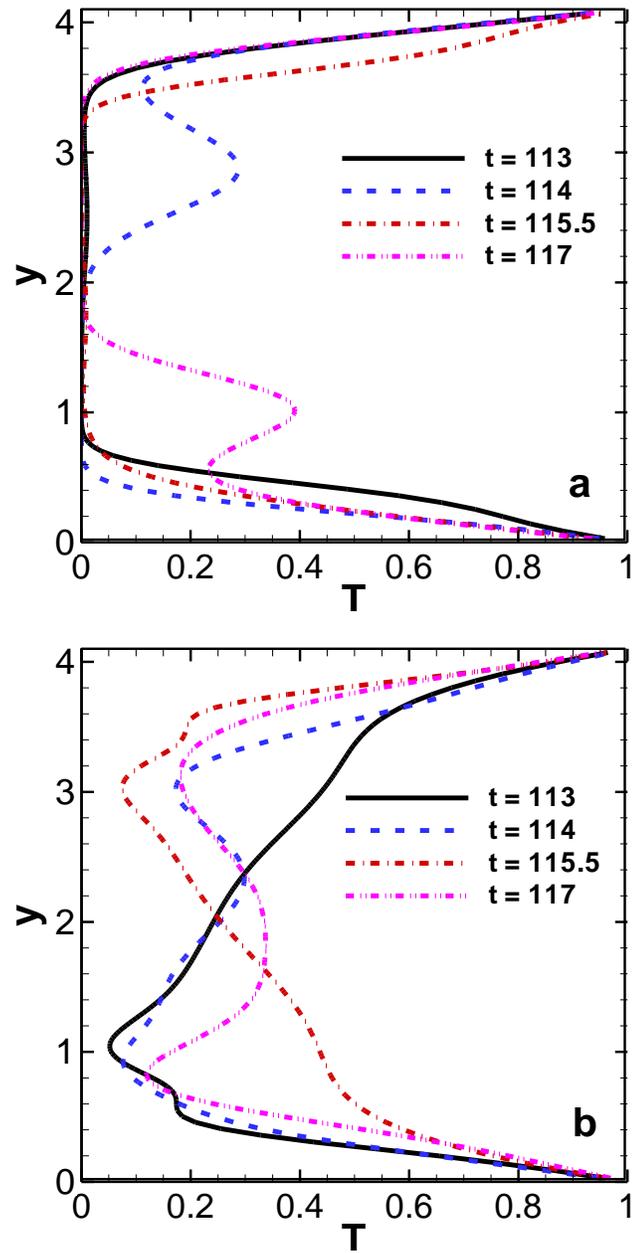

Fig. 11. Variation of temperature in a channel with a cylinder attached to an elastic plate along y-axis at $x =$ a) 8 and b) 27 for $Re = 100$, $Pr = 1$. The time instances are shown in the inset of Fig. 8 as black dots on time-varying position of the tip of the plate during a typical cycle of the plate oscillation.

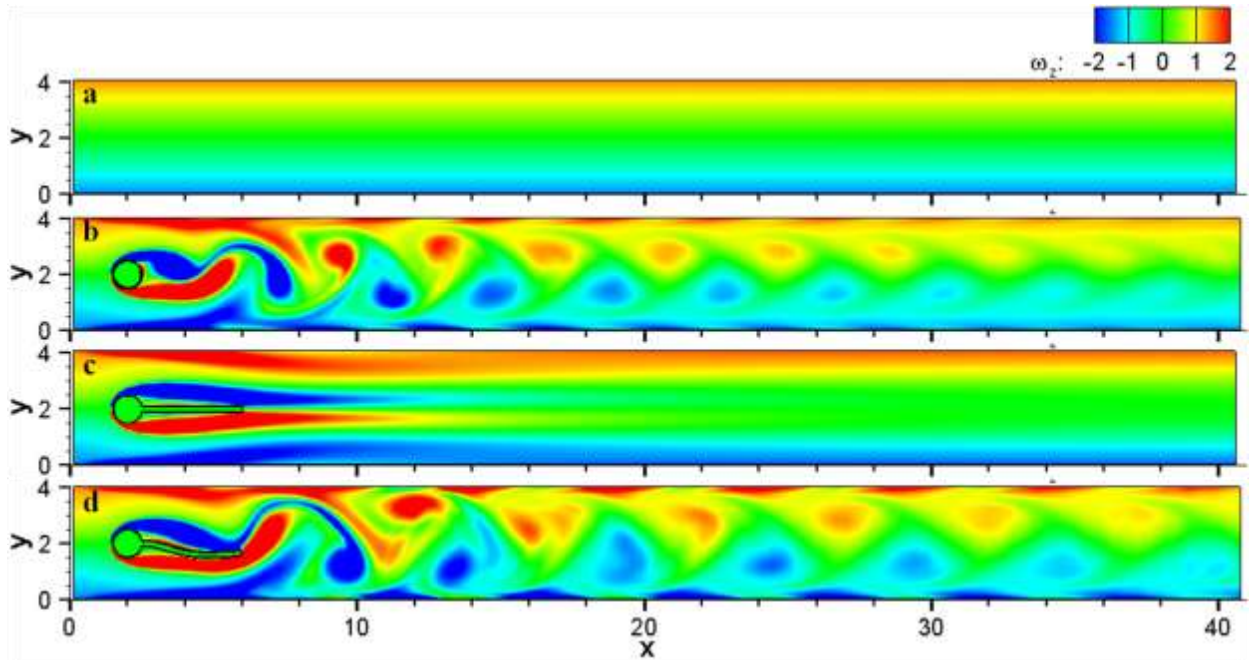

Fig. 12. Vorticity contours obtained at $t = 80$, for configurations CHL, CYL, CRP and CDP, defined in Table 1 at $Re = 100$.

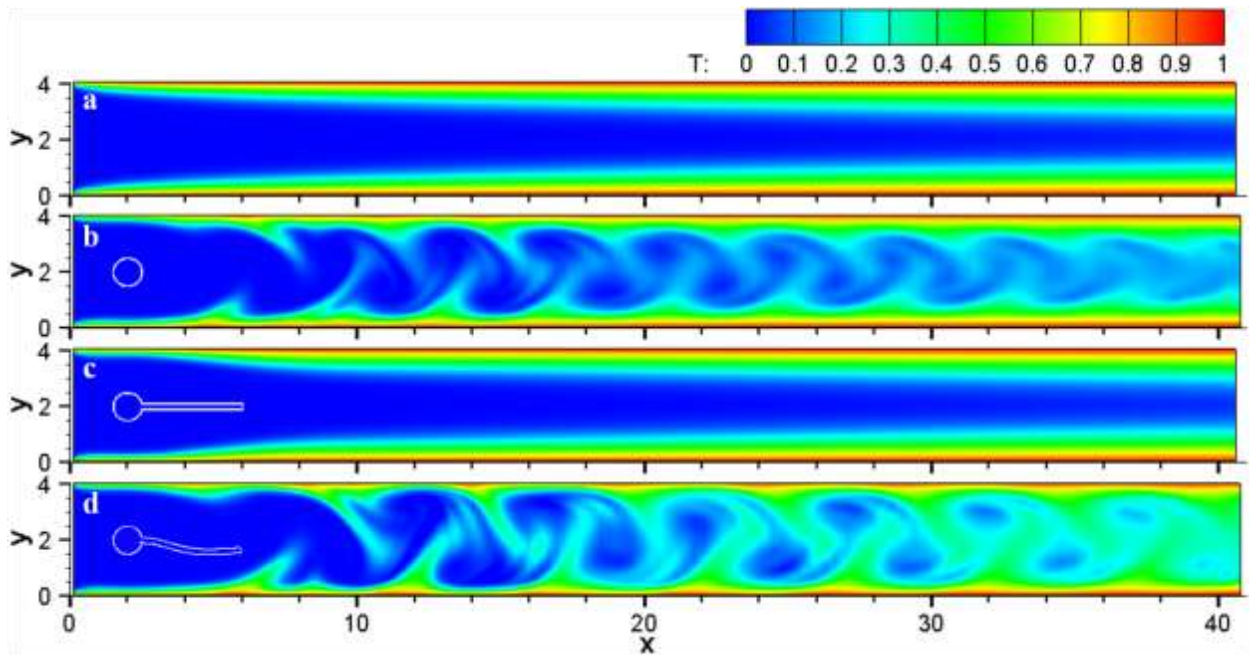

Fig. 13. Isotherms obtained at $t = 80$, for configurations CHL, CYL, CRP and CDP, defined in Table 1 at $Re = 100$, and $Pr = 1$.

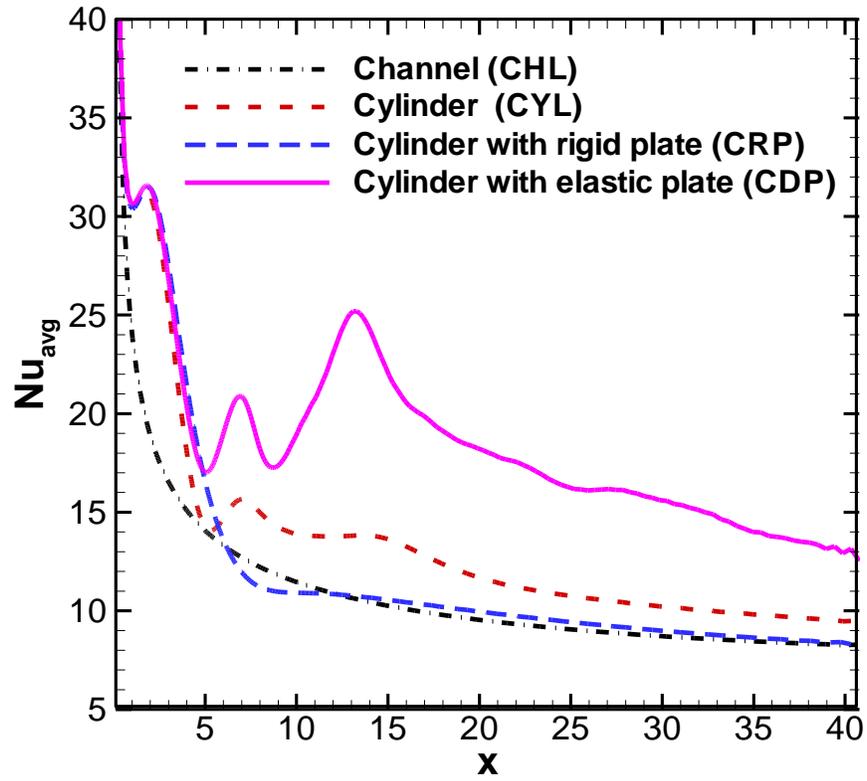

Fig. 14. Time-averaged Nusselt number ($Nu_{avg}$) for $Re = 100$ and $Pr = 1$, for configurations CHL, CYL, CRP and CDP, defined in Table 1.

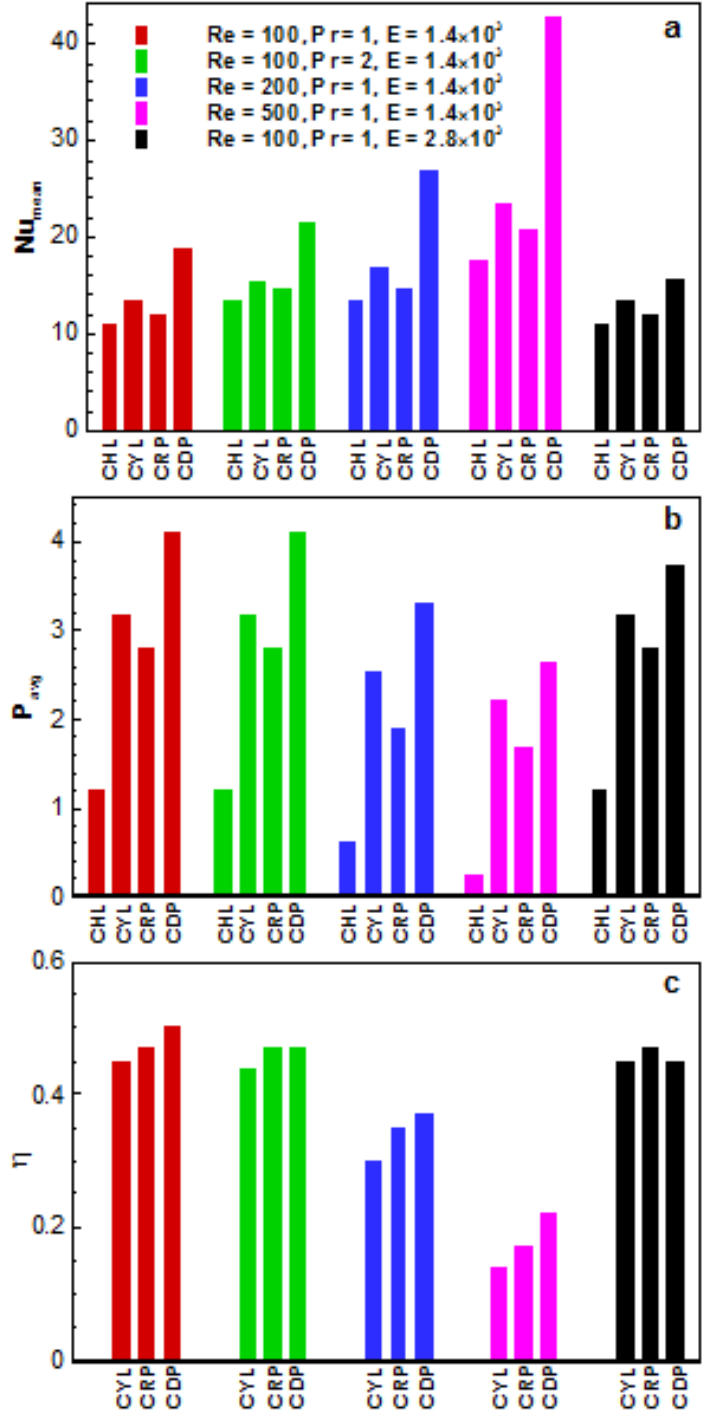

Fig. 15. (a) Time- and space-average Nusselt number, (b) Time-average pumping power, and (c) efficiency index for configurations CHL, CYL, CRP and CDP, defined in Table 1. The Reynolds number, Prandtl number and Young Modulus of the plate are varied in separate cases keeping other two parameters as constant.